# Gigahertz Clocked Quantum Key Distribution in Passive Optical Networks

Veronica Fernandez (1), Robert J. Collins (1), Karen J. Gordon (1),
Paul D. Townsend (2), and Gerald S. Buller (1)
(1) School of Engineering and Physical Sciences, Heriot-Watt University, Edinburgh, EH14 4AS, UK
(3) Tyndall National Institute and Department of Physics, University College Cork, Cork, Ireland

*Abstract*—Two multi-user approaches to fiber-based quantum key distribution systems operating at gigahertz clock frequencies are presented, both compatible with standard telecommunications fiber.

## I. INTRODUCTION

MOST current implementations of quantum key distribution (QKD) are *point-to-point* systems with one sender transmitting to only one receiver. Development of these single-receiver systems has now reached a comparatively advanced stage with clock rates in excess of 1GHz now achievable [1][2][3]. However, many communication systems operate in a *point-to-multi-point* (multiple-receiver) configuration rather than in *point-to-point* mode so it is crucial to demonstrate compatibility with this type of network in order to maximize the application range for QKD.

We have previously developed a point-to-point QKD system using the B92 protocol [4] with polarization encoding that operates at clock rates of up to 3.33 GHz over short

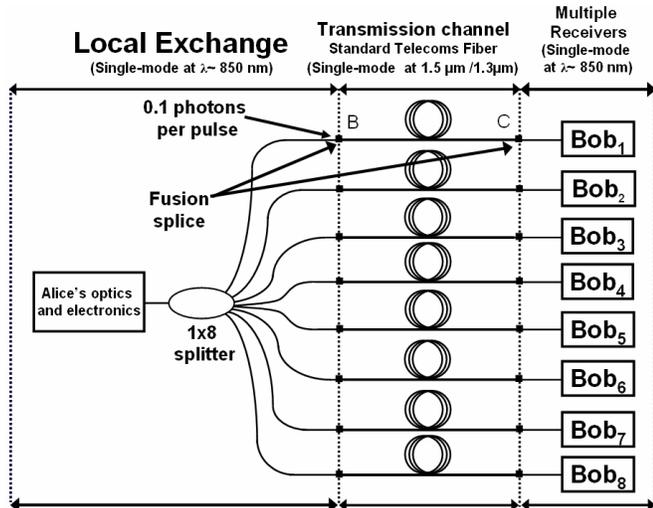

Fig.1. The passive optical splitter within the local exchange. Alice, the splitter and all of the Bobs are constructed from fiber which is single-mode at a wavelength of 850 nm. The transmission channel is standard telecommunications fiber.

This work was supported in part by the European Commission SECOQC Integrated Project and United Kingdom Engineering and Physical Sciences Research Council (project reference GR/N12466). Paul Townsend would like to thank Science Foundation Ireland for support under grant number 03/IN1/1340.

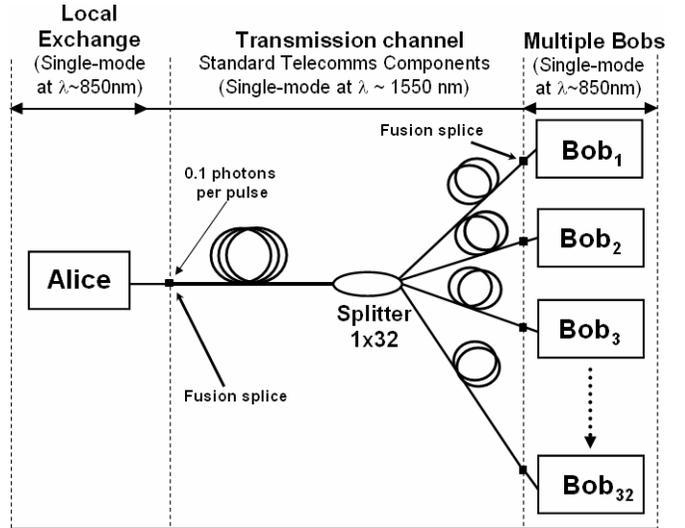

Fig. 2. The passive optical splitter within the insecure quantum transmission channel. Alice and all of the Bobs are constructed from fiber which is single-mode at a wavelength of 850 nm, the quantum channel is constructed from telecommunications fiber.

distances (< 20km) of standard telecommunications fiber at a wavelength of 850 nm using silicon single-photon avalanche diodes (SPADs) [1][3]. This system has now been extended to demonstrate two new passive optical network (PON) based point-to-multi-point QKD systems.

In the modern telecommunications industry PONs are currently deployed in optical access networks that link end customers to their nearest telecommunication provider's local exchange. There are a range of possible architectures for PONs characterized by the distribution and location of the passive optical splitters. Here we consider two representative cases which simulate the addition of QKD capability to existing, pre-deployed PONs. . In the first, the splitter is located within the local exchange and distributes data to each receiver via dedicated fiber links, as shown in Fig. 1. In the second the splitter is placed in the transmission channel, outside of the exchange as shown in Fig. 2

## II. QUANTUM KEY DISTRIBUTION NETWORKS

The point-to-point system of [1], and [3], provided the basis for the implementation of two different multi-user [5][6] applications. The first application utilized a single-mode at $\lambda \sim 850$ nm 1×8 splitter within the secure Alice station (as shown in Fig. 1). The use of a non-standard splitter is justified here because in practice the QKD system could employ a separate splitter to the conventional 1.3/1.5μm PON system, which is assumed to be operating over the same fiber distribution network (system separation being obtained through the use of wavelength division multiplexers). In contrast, in the second configuration the QKD system must use the same splitter as the conventional PON system, which is assumed to be pre-deployed. Hence a standard 1×32 device, which was single mode at 1.5 μm, was used within the insecure fiber transmission network (as shown in Fig. 2) Both applications were characterized in terms of transmission distance, net bit rate (after key distillation) and quantum bit error rate (QBER).

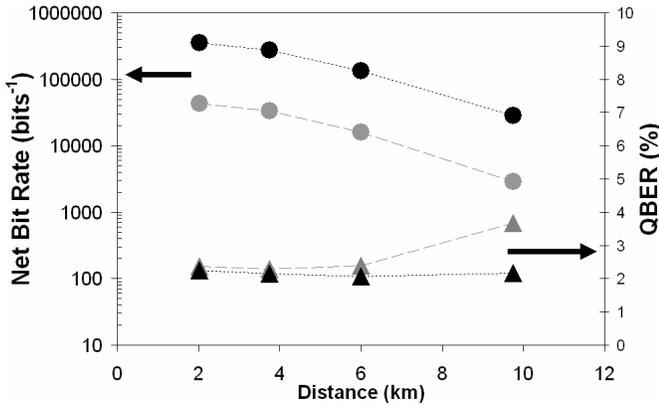

Fig. 3. Net Bit Rate and QBER versus fiber length for a clock frequency of 1.25 GHz for the splitter within Alice's secure station. These points were measured with a conventional μ of 0.1 photons per pulse at the output arms. The gray points denote projected values for an aggregate μ of 0.1 photons per pulse at the 8 output arms of the splitter

### A. Splitter within the Secure Alice Station

Fig. 3 shows the QBER and net bit rate at a clock frequency of 1.25 GHz obtained for four different transmission fiber distances (points B to C in Fig. 2). This application is equivalent to a point-to-point system between Alice and each Bob, but with a significant reduction in complexity and cost since only one transmitter is required for the network. In this experiment the attenuation was adjusted to give the conventional mean photon number ($\mu$) of ~0.1 photons per pulse at the output arms of the splitter.

We note, of course, that this particular approach gives an additional security disadvantage because a powerful eavesdropper with the capability to monitor all 8 output ports simultaneously would receive an aggregate mean photon number per pulse of up to ~0.8.

The gray points in Fig. 3 are projected values of net bit rate and QBER for an aggregate $\mu$ of 0.1 photons per pulse for all 8 fibers - the photon level where the above security disadvantage is removed.

### B. Splitter within the Insecure Quantum Channel

In this application, the splitter and its feeder fiber are situated in the external part of the network and are therefore potentially all fully accessible to an eavesdropper. As a consequence the mean photon number at the input to the feeder fiber must be maintained at a low level ($\mu \sim 0.1$ was employed in the experiments). Fig. 4 shows the net bit rate for the system for four different transmission channel lengths for a clock frequency of 1 GHz.

### III. CONCLUSION

Two different approaches to a multi-user QKD system were studied. Both situations are representative of the type of broadband PONs architectures that are currently being deployed in telecommunication access networks. Excellent performance is achieved for the QKD channel in all experiments, with QBER values in the range 2-5% and NBRs in the range 3000 to 40000 bits$^{-1}$, depending on configuration and distance.

### ACKNOWLEDGMENT

The authors gratefully acknowledge the assistance provided by Prof S. D. Cova and Dr I. Rech at the Politecnico di Milano in developing the detector used in this system.

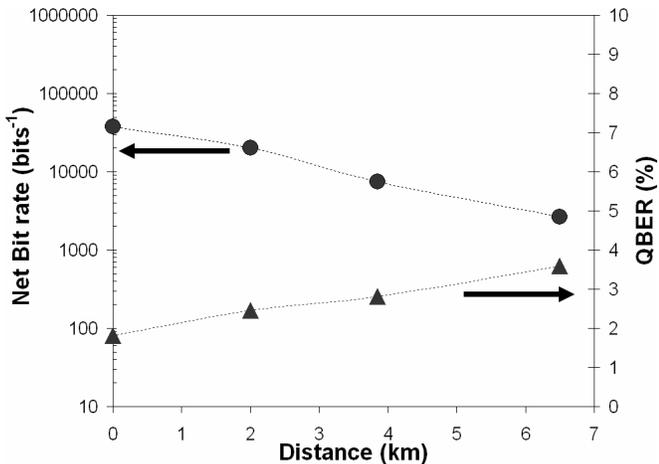

Fig. 4. Net Bit Rate and QBER versus fiber length for a clock frequency of 1 GHz for the splitter within the insecure fiber. These points were measured with μ set to 0.1 photons per pulse at the feeder fiber.